\documentclass[3p,times]{elsarticle}

\usepackage{graphicx}

\begin{document}

\begin{frontmatter}

\title{$I = 0$ $C = +1$ mesons from 1920 to 2410 MeV}

\author[c]{A.V.~Anisovich}
\author[a]{C.A.~Baker}
\author[a]{C.J.~Batty}

\author[b]{D.V.~Bugg\corref{cor1}}
\ead{david.bugg@stfc.ac.uk}

\author[b]{C.~Hodd}
\author[d]{H.C.~Lu}
\author[c]{V.A.~Nikonov}
\author[c]{A.V.~Sarantsev}
\author[c]{V.V.~Sarantsev}
\author[d]{B.S.~Zou}

\cortext[cor1]{Corresponding author}

\address[a]{Rutherford Appleton Laboratory, Chilton, Didcot OX11 0QX,UK}
\address[b]{Queen Mary and Westfield College, London E1\,4NS, UK}
\address[c]{PNPI, Gatchina, St. Petersburg district, 188350, Russia}
\address[d]{Institute of High Energy Physics, CAS, Beijing 100039, China}

\begin{abstract}
A combined fit is presented to data on $\bar pp$ annihilation in flight
to final states $\eta \pi ^0 \pi ^0$, $\pi ^0\pi ^0$, $\eta \eta $,
$\eta \eta '$ and $\pi ^-\pi ^+$.
The emphasis lies in improving an earlier study of $\eta \pi ^0 \pi ^0$ by
fitting data at nine $\bar p$ momenta simultaneously and with parameters
consistent with the two-body channels.
There is evidence for all of the $I = 0$, $C = +1$ $\bar qq$ states
expected in this mass range.
New resonances are reported with masses and widths ($M, \Gamma$) as follows:
$J^{PC} = 4^{-+}$ ($2328 \pm 38$, $\Gamma = 240 \pm 90$) MeV,
$1^{++}~(1971 \pm 15$, $240 \pm 45$) MeV,
$0^{-+}~(2285 \pm 20$, $325 \pm 30$)  MeV, and
$0^{-+}~(2010 ^{+35}_{-60}$, $270 \pm 60$)  MeV.
Errors on the masses and widths of  other resonances are also reduced
substantially.
All states lie close to parallel straight line trajectories of excitation
number v. mass squared.

\end{abstract}
\end{frontmatter}

Data from $\bar pp$ interactions in flight have the great merit of allowing a
direct study of $s$-channel meson resonances $R$ in formation reactions of the
type $\bar pp \to R \to A + B$.
Many decay channels $A$, $B$ may be studied.
Earlier, we have presented data on $\bar pp \to \eta \pi ^0 \pi ^0$ [1],
in which evidence was found for a number of  $I = 0$ $C = +1$ mesons.
That analysis fitted partial waves separately at each of nine $\bar p$
momenta;
resulting magnitudes and phases were then interpreted in terms of resonances.
[Table 1 shows the relation between beam momenta and centre of mass energies.]
Here our objective is to fit these data at all momenta simultaneously.
The fit to resonance parameters is also made simultaneously  to data on
$\bar pp \to \pi ^0 \pi ^0$, $\eta \eta$, $\eta \eta '$ and $\pi ^-\pi ^+$
reported earlier [2,3].

This combined fit is constrained strongly by the requirement of consistency
between channels and gives accurate determinations of
masses and widths of most resonances found in the data.
Fluctuations which were present in Refs. [1] at individual momenta are
eliminated; this is particularly important in restricting phases to a smooth
energy dependent behaviour consistent with analyticity.
In the earlier work, fluctuations in the large
high partial waves obscured details in
small amplitudes, particularly low partial waves.
We are now able to identify further $s$-channel resonances with
$J^{PC} = 4^{-+}$, $1^{++}$ and $0^{-+}$.
The essential results of the new analysis are summarised by the masses
and widths given in Table 2. These results supercede earlier determinations in
Refs. [1] and [3]. The present paper completes our analyses of channels
with $I = 0$, $C = +1$.
\begin{table} [htp]
\begin{center}
\begin{tabular}{cc}
Beam momentum & $\sqrt {s}$   \\
    (MeV/c)    & (MeV)         \\\hline
600 & 1962 \\
900 & 2049 \\
1050 & 2098 \\
1200 & 2149 \\
1350 & 2201 \\
1525 & 2263 \\
1642 & 2304 \\
1800 & 2360 \\
1940 & 2409 \\\hline
\end{tabular}
\caption {Momenta at which data are available and corresponding centre of
mass energies.}
\end{center}
\end{table}

We begin by outlining the considerations which influence the form of
our partial wave analysis.
The first point is that the well-known $f_4(2050)$ appears strongly in
all data sets, although we find a distinctly lower mass than that
quoted by the Particle Data Group (PDG) [4].
The $f_4(2050)$ acts as an interferomenter.
Its interferences determine relative phases of other partial waves.
These relative phases are found to vary little with mass, hence
requiring that all partial waves follow
a similar resonant behaviour to $f_4(2050)$.
The way the $\eta \pi \pi$ analysis goes is that one immediately finds strong
$J^{PC} = 3^{++}$ and $2^{-+}$ resonances.
The two-body data require $2^{++}$ and $0^{++}$ resonances.
Then, with increasing attention to detail, $4^{-+}$, $1^{++}$ and $0^{-+}$
resonances emerge from the $\eta \pi \pi$ data.

We therefore parametrise each partial wave amplitude as a sum of
resonances, plus a constant or slowly varying background where necessary.
In analysing earlier data [1,3], we have found that resonances mostly cluster
into two groups (a) from 1920 to 2050 MeV around $f_4(2050)$,
(b) from 2220 to 2320 MeV, around $f_4(2300)$.

The background terms may parametrise the tails of resonances below
the $\bar pp$ threshold or the effects of $t$-channel exchanges.
Singularities due to those exchanges are distant ($s \le 1$ GeV$^2$), so it
is to be expected that backgrounds will contribute mostly to low partial
waves. This is what we find.
The strong high partial waves with $J^{PC} = 4^{++}$  and $3^{++}$ are
consistent with no background.
In low partial waves ($J\le 2$), the background is parametrised as a broad
resonance (in most cases below the $\bar pp$ threshold) or as a constant.
Parametrising with resonances or constants guarantees that partial wave
amplitudes obey the important constraint of analyticity, since a
Breit-Wigner amplitude is analytic.
We see no evidence that strong threshold effects are present to perturb
such a parametrisation.

Each partial wave amplitude then takes the form:
\begin {equation}
f = \sum _i \frac {g_i\exp (i\phi _i)B_i(\bar pp)B_i(AB)}
          {M_i^2 - s - iM_i\Gamma _i}.
\end {equation}
The factors $B(\bar pp)$ and $B(AB)$ are standard Blatt-Weisskopf centrifugal
barrier factors which guarantee the correct threshold behaviour for each
channel; expressions are given in Ref. [5].
A common radius for the centrifugal barrier is fitted
to all partial waves.
There is a strong optimum at $0.829 \pm 0.021$ fm.
The full widths $\Gamma _i$ of all resonances are taken to be constant
because of the large number of open channels.
Each resonance is fitted with a real coupling constant $g_i$ and
a phase $\phi _i$.

We now discuss the channels fitted to the $\eta \pi \pi$ data.
Fig. 1 shows mass projections on to $\pi \pi$ and $\eta \pi$ at two
representative momenta; further figures are to be found in Refs. [1].
In Fig. 1, data are uncorrected for (small) variations of acceptance,
which are included in the maximum likelihood fit.
Histograms show the results of the present fit.
In Ref. [1], final states fitted to $\eta \pi ^0\pi ^0$  data were dominantly
$a_2(1320)\pi$, $f_2(1270)\eta$ and $\eta \sigma$, where $\sigma$ stands for
the $\pi \pi$ S-wave amplitude.
There were small additional contributions from
$a_0(980)\pi$, $f_0(980)\eta$ and $f_0(1500)\eta$.
In the $\eta \pi$ mass projection at the higher beam momenta there is a
distinct shoulder in the mass range around 1450 MeV.
A fit to this part of the the mass spectrum and higher $\eta \pi$ masses
requires further small contributions from $a_0(1450)\pi$ and
$a_2(1660)\pi$.
The evidence for $a_2(1660) \to \eta \pi$ from the Crystal Barrel experiment
is given in Ref. [6].
Parameters for $a_0(1450)$ are fixed to
values of the PDG.
However, evidence presented here for $s$-channel resonances does not depend
significantly on these small $a_2(1660)\pi$ or $a_0(1450)$ contributions.
We have tried including $f_0(1370)$ in the fit, but find negligible effect.
It is known to decay weakly to $\pi \pi$ [7] compared with $f_0(1500)$,
and the latter is already a small effect in the present data.

The 2-body channels $\pi \pi$, $\eta \eta$ and $\eta \eta '$ are
related by SU(3), since $\pi$, $\eta$ and $\eta '$ are members of a
single nonet.
Formulae for these constraints are given in Refs. [3].
In outline, SU(3) allows $\eta \eta$ and $\eta \eta '$ amplitudes to be
predicted from $\pi \pi$ using the well-known composition of $\eta $ and
$\eta '$ in terms of $(u\bar u + d\bar d)/\sqrt {2}$ and $s\bar s$.
Here we use the
assumption that $\bar pp$ does not couple directly to $\bar ss$;
this assumption has been tested in Refs. [3] and is well obeyed with
one striking exception.
The $f_0(2105)$ decays more strongly to $\eta \eta$ than to $\pi \pi$ by a
factor 1.88; this compares with the SU(3) prediction of $0.8^4 =0.42$.
That is, the decay to $\eta \eta$ is a factor 4.5 stronger than
predicted by the SU(3) relation.
Since the $f_0(2105)$ is produced strongly, this feature suggests
it has exotic character. Its coupling constants to
$\pi \pi$, $\eta \eta$ and $\eta \eta '$ are therefore fitted freely.

For $\eta \pi \pi$, amplitudes for $a_2(1320)\pi$ and $f_2(1270)\eta$
are again in principle related by SU(2) constraints if one assumes
ideal mixing for the $2^+$ nonet.
These have been tried in the fit, but do not work well.
This is probably because (a) there is a large mass difference between
these channels and (b) the $f_2(1270)\eta$ threshold is close to the
region we are analysing.
Consequently, momenta in these channels are
substantially different. This is likely to have large effects on
matrix elements. We find it necessary to fit magnitudes and phases of
$a_2(1320)\pi$ and $f_2(1270)\eta$ amplitudes freely.

Mass differences between $\pi $, $\eta$ and $\eta '$ also give rise to
differences in momenta $q$ in $\pi \pi$, $\eta \eta$ and $\eta \eta '$
channels, hence
significantly different form factors and centrifugal barrier effects.
We fit a form factor $\exp (-\alpha q^2)$ to all decays. The value of
$\alpha$ optimises at $0.92 \pm 0.10$ GeV$^{-2}$.

We find that two minor improvements may be made to the fit.
The phases $\phi _i$ of each resonance arise from multiple scattering in
initial and final states, as explained in Ref. [8].
Firstly, to allow for overlap of resonances (hence departure from
strict Breit-Wigner forms), phases $\phi _i$ are allowed to vary by up
to $\pm 20 ^{\circ }$ for $\eta \eta$ and $\eta \eta '$ with respect
to the $\pi \pi$ value.
Secondly, we find significant improvements by
allowing coupling constants $g_i$ to vary for $\eta \eta$ and $\eta \eta '$
from their SU(3) values by factors constrained to the range 0.7 to 1.3.
This probably reflects differences in form factors for different matrix
elements, which may be affected by masses, hence momenta, in the two-body
final state.
These refinements have only minor effects;
without them, conclusions on fitted resonance masses and widths change
by only a few MeV.

For $J^P = 2^+$ and $4^+$, $\bar pp$ may couple with orbital angular momentum
$L = J \pm 1$ (e.g. $^3P_2$ and $^3F_2$).
An ideal resonance should have the same phase for both $L$ values
(via multiple scattering through the resonance to each channel).
We therefore take the ratio of coupling constants $r_J =
g_{J+1}/g_{J-1}$ to be real. In fitting $\eta \pi \pi$
data, where a final state such as $f_2(1270)\eta$ may also have two or more
$L$ values,
the ratio of coupling constant is likewise constrained to be real in initial
fits.
Again, we find small but significant improvements to the fit if these
ratios for $f_2\eta$ and $a_2\pi$ decays are allowed to depart from
real values by phase angles in the range
up to $\pm 20 ^{\circ }$. But there is little effect on fitted masses and
widths, merely a small improvement in the quality of the fit to data.

In Refs. [3], data on $\pi \pi$, $\eta \eta$ and $\eta \eta '$ required four
$J^{PC} = 2^{++}$ resonances.
In addition, data on $\bar pp \to (\eta \eta )\pi ^0$ provide evidence for a
broad $2^+~\eta \eta$ contribution with $M = 1980 \pm 50$, $\Gamma = 500 \pm
100$ MeV [9].
Central production of $4\pi$ provides even clearer evidence with very similar
mass and width [10].
We find that this broad component improves the present combined fit to $\eta
\pi \pi$ and 2-body channels strongly.
Here we find a clear optimum for the mass at
$M = 2010 \pm 40 $ MeV and for the width at $495 \pm 50$ MeV.
These values are consistent with Refs. [9] and [10].
However, those earlier determinations have the advantage that the broad $2^+$
component is clearly visible by eye, hence providing the incentive for
trying it here.
Without this component, large interferences develop between the four $2^+$
states, so as to simulate this broad contribution.

\begin{table} [htp]
\begin{center}
\begin{tabular}{cccccccc}
I & $J^{PC}$ & Mass $M$ & Width $\Gamma$ & $r$ & $\Delta S (\eta \pi \pi )$ & \\
  &        & (MeV)    & (MeV)          &     & &  \\\hline
0 & $6^{++}$ & $2485 \pm 40 $ & $410 \pm 90$ & 0 & 245 \\
& $4^{++}$ & $2283 \pm 17 $ & $310 \pm 25$ & $2.7 \pm 0.5$ & 2185 \\
& $4^{++}$ & $2018 \pm 6 $ & $182 \pm 7$ & $0.0 \pm 0.04$ & 1607 \\
& $4^{-+}$ & $2328 \pm 38 $ & $240 \pm 90$ & & 558 & New\\        \
& $3^{++}$ & $2303 \pm 15$ & $214 \pm 29$ & & 1173 \\
& $3^{++}$ & $2048 \pm 8 $ & $213 \pm 34$ & & 1345\\
& $2^{++}$ & $2293 \pm 13 $ & $216 \pm 37$ & $-2.2 \pm 0.6$ & 1557 \\
& $2^{++}$ & $2240 \pm 15 $ & $241 \pm 30$ & $0.46 \pm 0.09$ & 468\\
& $2^{++}$ & $2001 \pm 10 $ & $312 \pm 32$ & $5.0 \pm 0.5$ & 168 \\
& $2^{++}$ & $1934 \pm 20 $ & $271 \pm 25$ & $0.0 \pm 0.08$ & 1462\\
& $2^{++}$ & $2010 \pm 25 $ & $495 \pm 35$ & $1.51 \pm 0.09$ & 694 \\
& $2^{-+}$ & $2267 \pm 14 $ & $290 \pm 50$ & & 1349 \\
& $2^{-+}$ & (2030)         & (205)        &  \\
& $2^{-+}$ & (1860)         & (250)        &  \\
& $1^{++}$ & $2310 \pm 60 $ & $255 \pm 70$ & & 882 \\
& $1^{++}$ & $1971 \pm 15 $ & $240 \pm 45$ & & 1451 & New\\
& $0^{++}$ & $2337 \pm 14 $ & $217 \pm 33$ & \\
& $0^{++}$ & $2102 \pm 13 $ & $211 \pm 29$ & \\
& $0^{++}$ & $2040 \pm 38 $ & $405 \pm 40$ & \\
& $0^{-+}$ & $2285 \pm 20 $ & $325 \pm 30$ & & 1342 & New \\
& $0^{-+}$ & $2010 ^{+35}_{-60}$ & $270 \pm 60$ & & 1189 & New \\\hline
1 & $5^{--}$ & $2295 \pm 30 $ & $235 ^{+65}_{-40}$ & 0 \\
& $3^{--}$ & $2300 ^{+50}_{-80} $ & $340 \pm 50$ & $2.00 \pm 1.1$ \\
& $3^{--}$ & $2210 \pm 40 $ & $360 \pm 50$ & $-0.08 \pm 0.23$ \\
& $3^{--}$ & $1981 \pm 14 $ & $180 \pm 35$ & $0.006 \pm 0.024$ \\
& $1^{--}$ & $2145 \pm 50$ & $270 \pm 100$ & $2.03 \pm 1.25$\\
& $1^{--}$ & $2000 \pm 30 $ & $295 \pm 85$ & $1.05 \pm 0.24$ \\\hline
\end{tabular}
\caption {Resonances fitted to the data. Errors cover systematic
variations observed in a variety of fits with different ingredients.
Values in parentheses are fixed from other data [18].
The sixth column shows changes in $S = $ log likelihood when each component
is omitted in turn from the fit to $\eta \pi ^0 \pi ^0$ data and remaining
components are re-optimised.
The final column highlights those resonances which are new.}
\end{center}
\end{table}

We have searched for ambiguous solutions by (i) removing each resonance (or
background) one by one and re-optimising the rest, (ii) changing signs of
amplitude ratios $r$ one by one and
(iii) moving resonances in steps of 20 MeV in mass and
40 MeV in width over large ranges (9 steps each).
Each iteration of the overall fit takes 10 minutes of
computing and one solution converges in typically 20--50 iterations. So
$\sim 500$ alternatives have been explored. All variants collapse
back to the same solution.

Intensities of all amplitudes fitted to $\eta \pi ^0 \pi ^0$ are displayed in
Fig. 2.  They have been corrected for all $\eta$ and $\pi ^0$ decays, i.e.
for the 39.25\% branching fraction of
$\eta \to \gamma \gamma$ and the 98.798\% branching fraction of
$\pi ^0 \to \gamma \gamma$.
Typical errors are $\pm 5\%-10\%$ for the larger partial waves,
increasing to $\pm 35\%$ for the smallest.
Amplitudes which contribute less than 0.3\% of the integrated cross
section are found to change log likelihood by $<40$ and are omitted.Table 2 shows masses and widths of fitted resonances.
Statistical errors are negligible, so quoted errors cover the range of
solutions observed when components of the fit are varied.
It is apparent, for example, that there are small systematic discrepancies
between data on final states $\pi ^-\pi ^+$ and $\pi ^0 \pi ^0$, and
some masses and widths change by a few MeV according to the way one weights
different data sets.
Errors include systematic changes when the radius of the centrifugal barrier is
varied within its error.
The last column of the Table shows changes in log likelihood when each
component is removed from the fit to $\eta \pi ^0 \pi ^0$ data and others
are re-optimised.
Our definition of log likelihood is such that it changes by 0.5 when one
parameter changes by one standard deviation.
For $2^+$ and $4^+$ components, there are also large changes in $\chi ^2$
arising in the fit to two-body data;
those changes are close to values quoted in Refs. [3].
Consequently, all of the components in the fit are highly significant.
The last 6 lines of Table 1 show, for completeness, parameters of
$I = 1$ resonance fitted to two-body data;
there are small changes from values of Refs. [3].

The fit to $\pi \pi$, $\eta \eta$ and $\eta \eta '$ has changed little from
that of Refs. [3], so attention will be concentrated here
on fits to $\eta \pi \pi$.
For those data there are some significant changes from the fit described in
Refs. [1]. The main origin of these changes is that ratios $r$ of amplitudes
with $L = J \pm 1$ are defined well by the two-body data, specifically
by the polarisation data on $\bar pp \to \pi ^- \pi ^+$. Using these
ratios in the combined fit introduces significant changes to $2^+$
amplitudes in $\eta \pi \pi$. A knock-on effect is that there is
also a large improvement in the fitted $3^+$ amplitude towards the top of the
mass range.

Argand diagrams are displayed in Fig. 3.
Crosses mark individual beam momenta;
Table 1 lists these beam momenta and the corresponding centre of mass
energies.
In many cases, one discerns in Fig. 3 that the amplitude varies rapidly in the
low mass region and again in the high mass region. This indicates the
presence of two resonances. However, only in Figs. 3(h) and (e) does one
see conspicuously separate loops due to different resonances.

The data require a larger and more linear phase variation than can be
produced by a single resonance. A sequence of
resonances conspires to produce approximately a linear phase variation with
mass squared $s$.
In electronics, it is well known that a linear phase variation with
frequency may be obtained with a Bessel filter [11], which has a sequence
of poles almost linearly spaced in frequency.
Meson resonances appear to behave in an analogous way as a function of
$s = M^2$.
Physically, such a system generates a time delay which is independent of
frequency, since group velocity is proportional to the gradient of phase v.
frequency.

We shall now comment on each $J^P$ in turn, beginning with the high partial
waves which are most conspicuous.
For $4^+$, the different peaks in Figs. 2(g) and (h) for $f_2(1270)\eta$
(full curves) and $a_2(1320)\pi $ (dashed) obviously
demand two resonances.
The lower one optimises at a mass of $2018 \pm 6$ MeV,
considerably lower than the PDG average of $2044 \pm 11$ MeV.
The effect of the centrifugal barrier is large and the intensity of
the $^3F_4$ partial wave peaks at 2080 MeV in Fig. 2(g).
Differences in earlier determinations of the mass may well depend on
varying treatments of the centrifugal barrier.
Our mass is determined essentially by the maximum in the `speed plot', i.e.
in the movement of the amplitude in the Argand diagram; it is found to be
quite insensitive to the radius chosen for the centrifugal barrier.
As the radius increases, all masses go up; however, with our error for
the radius of the barrier, the contribution to the uncertainty in  mass
is only 0.7 MeV. The width of the resonance does, however, correlate much
more strongly with the radius, and this contributes an uncertainty to the
width of 3.3 MeV.
The $L = 3$ centrifugal barrier in the $\bar pp$ channel suppresses strongly
the coupling of the lower $4^+$ resonance to $f_2(1270)\eta$.
The lower resonance couples purely to $\bar pp$ $^3F_4$, while the upper
one requires a large $^3H_4$ component.

For $J^P = 3^+$, the different peaks in Fig. 2(d) for $f_2\eta$ (full curves)
and $a_2\pi$ (dashed) again clearly demand two resonances.
The Argand diagram of Fig. 3(h) also clearly demands two resonances.
The upper resonance is considerably stronger in the present fit than in Refs.
[1].

For $J^P = 2^+$, the $\eta \pi \pi $ data alone do not resolve $f_2(1934)$ and
$f_2(2001)$ clearly, because both lie at the bottom end of
the available mass range. One sees from Table 2 that removing the
$f_2(2001)$ produces a change in log likelihood in $\eta \pi \pi$ data alone
of only 168. This is
because it may be interchanged to some extent
with contributions from $f_2(1934)$.
For $f_2(1934)$ and $f_2(2001)$, the two-body data play a very important
role, for two reasons.
Firstly there are extensive data on the $\pi ^-\pi ^+$ channel down to
360 MeV/c (a mass of 1910 MeV), which determine quite well the mass and
width of the $f_2(1934)$.
This state is consistent in mass with the
the $f_2(1920)$ of GAMS [12] and VES [13], though they find a narrower width.
Secondly, the polarisation data
provide a sensitive determination of the  ratios of amplitudes between
$^3F_2$ and $^3P_2$.
The values of $r_2$ given in Table 2 are for ratios of coupling constants
$^3F_2/^3P_2$, i.e. after factoring out the effect of the centrifugal
barriers. Both $f_2(1934)$ and $f_2(2001)$ are definitely required by the
two-body data. If either of them is removed from the fit, $\chi ^2$
for 2-body data increases by $>2000$, as reported in Refs. [3].

The earlier analysis of $\eta \pi \pi$ data [1] showed the requirement for
three $2^+$ states at $2020 \pm 50$, $2240 \pm 40$ and $2370 \pm 50$ MeV.
These masses now adjust naturally by small amounts to those of Table 2.
The lowest $2^+$ state at 1934 MeV contributes strongly to
$\bar pp \to \pi \pi$ and is dominantly $^3P_2$; that at 2001 MeV is
largely $^3F_2$ and is close in mass to $^3F_3$ and $^3F_4$ resonances.
The $f_2(2240)$ and $f_2(2293)$ are dominantly $^3P_2$ and $^3F_2$
respectively.
Each of the Argand diagrams of Figs. 3(k) to (o) show at most a requirement
for two $2^+$ states.
However, the two-body data require four $2^+$ states, as reported in
Refs. [3].
The $f_2(1934)$ lies below the available mass range for $\eta \pi \pi$ data.

We have speculated earlier on the possibility that the broad $f_2(1980)$
has a large component due to the $2^+$ glueball [14,15].
In this context, its relative coupling to $\pi \pi$ and $\eta \eta$ is
important.
We fit these freely, with the result $g^2_{\eta \eta}/g^2_{\pi ^0 \pi ^0 } =
0.72 \pm 0.06$.
This compares with the prediction 0.41 for
$(u\bar u + d\bar d)/\sqrt {2}$ and
1 for a glueball.
The result lies midway between the two.
In Ref. [3], the amplitude was expressed in terms of flavour mixing to a
linear combination of states $\cos \Phi |u\bar u + d\bar d>/\sqrt {2} +
\sin \Phi |s\bar s>$. With this parametrisation, the flavour mixing angle
$\Phi $ optimises at $23.6 \pm 3.5^{\circ}$ compared with the value
$35.6^{\circ }$ for a glueball and $0^{\circ }$ for $(u\bar u + d\bar d)$.
Some mixing between a glueball and neighbouring $q\bar q$ states is likely.

For $J^P =0^+$, there are only minor changes from Ref. [3].
The $f_0(2105)$ makes a large and very well defined contribution to
$\eta \eta$, and a small contribution to $\pi \pi$. It requires a
mixing angle $\Phi = (58 \pm 5)^{\circ }$. Its strong production and
large flavour mixing angle suggests exotic character. It is clearly an
`extra', non-$q\bar q$  state.
Its mass optimises at $2102 \pm 13$ MeV and its width at $211 \pm 29$ MeV.
These values agree closely with those found in the E760 experiment [16].
Since their determinations of the width is somewhat more precise, namely
$203 \pm 10$ MeV, we use this value in the final fit.
The data also require the presence of another
nearby broad $f_0(2040)$, in good agreement with WA102 parameters [17]
and  a further $f_0(2337)$ at higher mass.

For $J^{PC} = 2^{-+}$, earlier data on $\bar pp \to \eta \pi ^0 \pi ^0 \pi ^0$
demonstrate the presence of two states at 1860 and 2030 MeV
[18].
The first of these decays largely to $[f_2\eta ]_{L = 0}$ and
the second
decays weakly to this channel. The present $\eta \pi \pi$ data extend only down
to 1960 MeV and therefore do not resolve these two states.
We therefore fix their masses, widths and branching ratios between
$f_2\eta$, $a_2\pi$ and $a_0\pi$ channels to the values of Ref. [18].
The $[f_2\eta ]_{L = 0}$ channel
makes a dominant contribution to the $2^{-+}~\eta \pi \pi$ partial waves at low
mass, see Fig. 2(c), full curve.
The presence of a further $\eta _2$ at high masses is seen most clearly
in $^1D_2 \to [f_2\eta ]_{L = 2}$, Fig. 2(c) chain curve; it is also visible in
the smaller components
$[a_0(980)\pi ]_{L = 2} $, Fig. 2(k) dashed curve and in
$^1D_2 \to [f_0(1500)\eta ]_{L = 2}$, Fig. 2(k) chain curve.
Its mass, $2267 \pm 14$ MeV, is somewhat lower than the earlier determination
[1], $2300 \pm 40$ MeV.
However, it agrees closely with a conspicuous peak
observed [19] in the integrated cross section for $\bar pp \to \eta '\pi ^0\pi
^0$. There, a mass of $2248 \pm 20$ MeV is found.
The width from present data is $290 \pm 50$ MeV.
However, the width is determined better by $\eta '\pi ^0 \pi ^0$ data:
$280 \pm 20$ MeV, and we fix it at this value in the final fit to present data.

We now come to the new states.
There is a well known $\rho _5$ ($\bar qq~^3G_5$)  listed by the PDG at
2330 MeV. One expects a
$q\bar q$ $4^{-+}$ state ($\bar qq~^1G_4$) of similar mass.
We now observe small but well determined contributions from a $4^{-+}$
state at $2328 \pm 38$ MeV in
$^1G_4 \to [a_2\pi ]_{L = 2}$, $[a_2\pi ]_{L = 4}$ and $[a_0\pi ]_{L = 4}$,
Fig. 2(m).
This resonance, although it contributes the smallest change in
log likelihood in Table 2, is very stable throughout all fits.
There is no doubt of its presence, despite the fact that it contributes only
2\% of the integrated cross section.
However, there is quite a large error on the width.

Next we consider $1^{++}$.
The earlier work of Ref. [1] found a strong $[a_2\pi ]_{L = 1}$ intensity
near the $\bar pp$ threshold.
In the new combined fit,
the $1^{++}$ amplitude is large in several channels around 1975 MeV,
falling rapidly at higher masses, see Fig. 2(j).
Despite the fact that it lies at the bottom end of the available mass range,
parameters of the resonance in this range are very stable:
$M = 1971 \pm 15$ MeV, $\Gamma = 241 \pm 45$ MeV.
Argand diagrams of Figs. 3(c) and (d) for $[f_2\eta ]_{L = 1}$ and
$[\sigma \eta ]_{L = 1}$ require rapidly varying phases at low mass when one
remembers that the amplitude goes to zero at the $\bar pp$ threshold;
this rapid phase variation is a clear signature of the resonance.

At high masses, the phase variation observed in Refs. [1] suggested a
$1^{++}$ resonance at $2340 \pm 40$ MeV with $\Gamma = 340 \pm 40$ MeV.
We now find that this second $1^{++}$ resonance is definitely required.
Without it, log likelihood is worse by 882; statistically this is a
25 standard deviation effect when one allows for the number of fitted
parameters. Nonetheless, the mass of this state is the least
well determined of all the resonances: $M = 2310 \pm 60$ MeV.
The reason is that it makes small contributions
to the distinctive channels $a_2(1320)\pi$ and $f_2(1270)\eta$,
and there are large interferences with the $0^-$ amplitude.
As the resonance mass is changed, these interferences are able to
absorb the variation with small changes in log likelihood.
This resonance is most clearly visible in $[f_2 \eta ]_{L = 1}$,
Fig. 2(b), full curve.

Lastly we consider $0^{-+}$.
Here there is a large change, compared with Refs. [1], in the fit to the
broad background in the Dalitz plot.
Earlier this was fitted purely with the $\eta \sigma$ channel, although
it was evident in Refs. [1] that the $\eta \pi$ mass spectrum was not
fitted perfectly around a mass of 1450-1650 MeV. For that reason,
conclusions were not drawn earlier about structure in the $0^-$ partial wave.
This feature has now been improved substantially by the addition of
channels $a_0(1450)\pi ^0$ and $a_2(1660)\pi ^0$.

The $\sigma \eta$ and $f_0(1500)\eta$ channels, Fig. 2(i) (full and dotted
curves) now definitely require the presence of a high mass
resonance with $M = 2285 \pm 20$ MeV.
In addition, there is a very strong contribution close to the $\bar pp$
threshold in several channels of Figs. 2(a) and (i).
A resonance is definitely required in this mass region.
Without it, log likelihood gets worse by a very large amount, 1189.
The effect of dropping it is illustrated by the dashed curves of Figs.
3(a) and (b).
For this partial wave, amplitudes do not go to zero at the $\bar pp$
threshold, but instead go to values described by scattering lengths.
It is not possible to exclude the possibility that the
$\bar pp$ threshold plays a strong role (a cusp at the $\bar pp$ threshold).
This makes the identification of the mass of the lower resonance difficult.
The optimum is at $2010 ^{+35}_{-60}$ MeV.

There is just one significant change in the fit to two-body data, compared
with Refs. [3]. There is a strong $3^{--}$ resonance in the low mass range.
It interferes strongly with $2^{++}$ states in data on
$\bar pp \to \pi ^- \pi ^+$. Small  changes in the masses and widths fitted
to the $2^+$ states have had the effect of increasing the mass of this
$\rho _3$ from $1960 \pm 15$  MeV [3] to $1981 \pm 14$ MeV; the fitted width
has also increased slightly.

Fig. 4 illustrates the quality of the fit at two momenta.
It shows the angular distributions for production
of mass regions centred on (a) and (d) $f_2(1270)$,
(b) and (e) $a_2(1320)$,  (c) and (f) $a_0(980)$.
In all cases, there are background contributions underneath these resonances,
particularly in the case of the small $a_0(980)$ signal; these backgrounds
are included in Fig. 4.
The angular distributions are well reproduced by the fit in all cases.

We now turn to the interpretation of the results.
Fig. 5 shows plots of mass squared v. excitation for all $J^P$.
In making this plot, we use the K-matrix mass of 1598 MeV for $f_2(1565)$ [20],
and 1400 MeV for $f_0(1370)$ [7].
The reason for this choice is that the well known linear mass relation
between $\Delta (1230)$, $\Sigma (1385)$, $\Xi(1520)$ and $\Omega ^-(1672)$
works well when one uses for the $\Delta$ the mass at which the phase shift
goes through 90$^{\circ}$, i.e. the K-matrix mass.
It does not work nearly so well using the T-matrix pole position of 1210
MeV.
For Breit-Wigner resonances of constant width (used here), the K-matrix mass
and the T-matrix pole position are the same.

A remarkably simple pattern is apparent in Fig. 5.
All states lie close to parallel straight-line trajectories.
These trajectories extrapolate well to known states in the mass range
1200--1700 MeV.
[We do not, however, attempt to place the $\eta$ on the $0^-$ trajectory,
since its mass is affected strongly by the instanton interaction].
The simplicity of Fig. 5 suggests that observed states are $q\bar q$
rather than hybrids.
For $^3P_1$ and $^1S_0$, states expected around 1670 MeV are presently
missing.
For quantum numbers $^3D_3$, $I = 1$ resonances observed in the $\pi ^-\pi ^+$
channel are shown; it is worthwhile
to show these results, since the lowest two states are very well defined.

Using PDG masses for resonances below 1.9 GeV, slopes for different
quantum numbers are shown in Table 3. They are consistent with the
same slope within statistical errors; the mean slope is
$1.143 \pm 0.013$ GeV$^2$ per excitation.
The most accurate determination of the slope comes from the $^3P_2$
trajectory, which begins with the well known $f_2(1270)$.
However, one must expect some deviations from straight lines due to
local perturbations, for example (a) from mixing with nearby glueballs or
hybrids, (b) from level repulsion between $2^+$ states,
(c) from nearby thresholds, particularly in the low mass region,
and (d) from variation with $s$ of the effects of tensor and
spin-orbit splitting.
There is an `extra' $2^{-+}$ state at 1860 MeV, which will perturb
the $2^{-+}$ tratectory; as discussed in Ref. [18], it is a candidate for a
hybrid expected in that mass region.

The $0^+$ trajectory is drawn with $n = 1$ for $f_0(1370)$.
The line does, however, go through $f_0(980)$, and it is possible that
this state is the $n = 1$ ground state rather than a molecule.
The error on the K-matrix mass of $f_0(1370)$ is quite large, so the
$0^+$ trajectory is defined mostly by $f_0(1770)$ and $f_0(2337)$.
Because of strong mixing between scalar states, the interpretation of the mass
range around $f_0(980)$ and $f_0(1370)$ may be complicated; it has previously
been considered in terms of the K-matrix approach by Anisovich et al. [21].

\begin{table} [htp]
\begin{center}
\begin{tabular}{cc}
$J^P$ &  Slope   \\
     & (GeV$^2$)         \\\hline
$4^{++}$ & $1.139 \pm 0.037$\\
$3^{++}$ & $1.107 \pm 0.078$\\
$^3F_2 $ & $1.253 \pm 0.066$\\
$^3P_2 $ & $1.131 \pm 0.024$\\
$2^{-+}$ & $1.217 \pm 0.055$\\
$1^{++}$ & $1.120 \pm 0.027$\\
$0^{++}$ & $1.164 \pm 0.041$\\
$0^{-+}$ & $1.183 \pm 0.028$\\
$3^{--}$ $(I=1)$ & $1.099 \pm 0.029$\\\hline
\end{tabular}
\caption {Slopes of trajectories for different quantum numbers.}
\end{center}
\end{table}

In the absence of tensor and spin-orbit splitting,
$^3F_4$, $^3F_3$ and $^3F_2$ states are degenerate in mass.
Tensor and spin-orbit splitting may be assessed from the relations [22]
\begin {eqnarray}
\Delta M_{LS} &=& [-20M(^3F_2) - 7M(^3F_3) + 27M(^3F_4)]/54, \\
\Delta M_{T} &=& [-4M(^3F_2) + 7M(^3F_3) - 3M(^3F_4)]/14.
\end {eqnarray}
For both multiplets of Table 2, $^3F_3$ states lie highest in mass, requiring
significant tensor splitting.
This splitting is in the sense predicted by one-gluon exchange.
Using a linear confining potential plus one-gluon exchange, Godfrey and Isgur
[23] predict $\delta M_T = 9$ MeV for the lower multiplet and $\Delta M_{LS}
 = -20$ MeV.
From Table 2, one finds $\delta M_T = 20 \pm 4.5$ MeV for the lower
multiplet and $7.1 \pm 7.3 $ MeV for the upper one; errors take into account
observed correlations in fitted masses.
Because we use the same centrifugal
barrier for $^3F_4$, $^3F_3$ and $^3F_2$ states, there is almost no
correlation of $\Delta M_T$ or $\Delta M_{LS}$ with the radius of the barrier.
The observed spin-orbit splittings are $\delta M_{LS} = 2.4 \pm 4.4$ MeV
for the lower multiplet and $-6.3  \pm 5.1$ MeV for the upper one;
these average approximately to zero.

In summary, the expected $q\bar q$ states in this mass range are observed with
well determined masses and widths, except for the lower $0^-$ and the upper
$1^+$ state, where errors are sizeable.
They follow a simple pattern requiring approximately
linear trajectories of mass squared
against excitation number, with a slope of $1.143 \pm 0.013$ GeV$^2$
per excitation.
In addition, there is evidence for a broad $2^+$
component with $M  = 2010 \pm 25$ MeV, $\Gamma = 495 \pm 35$ MeV.
Tensor splitting for $^3F$ states is, within sizeable errors, consistent
with that predicted from one-gluon exchange and a linear confining potential.
Spin-orbit splitting of $^3F$ states is consistent with zero.

\section{Acknowledgement}
We thank the Crystal Barrel Collaboration for
allowing use of the data.
We acknowledge financial support from the British Particle Physics and
Astronomy Research Council (PPARC).
We wish to thank Prof. V. V. Anisovich for helpful discussions.
The St. Petersburg group wishes to acknowledge financial support from PPARC and
INTAS grant RFBR 95-0267.


\begin{figure}
\centerline{\includegraphics[width=0.8\textwidth]{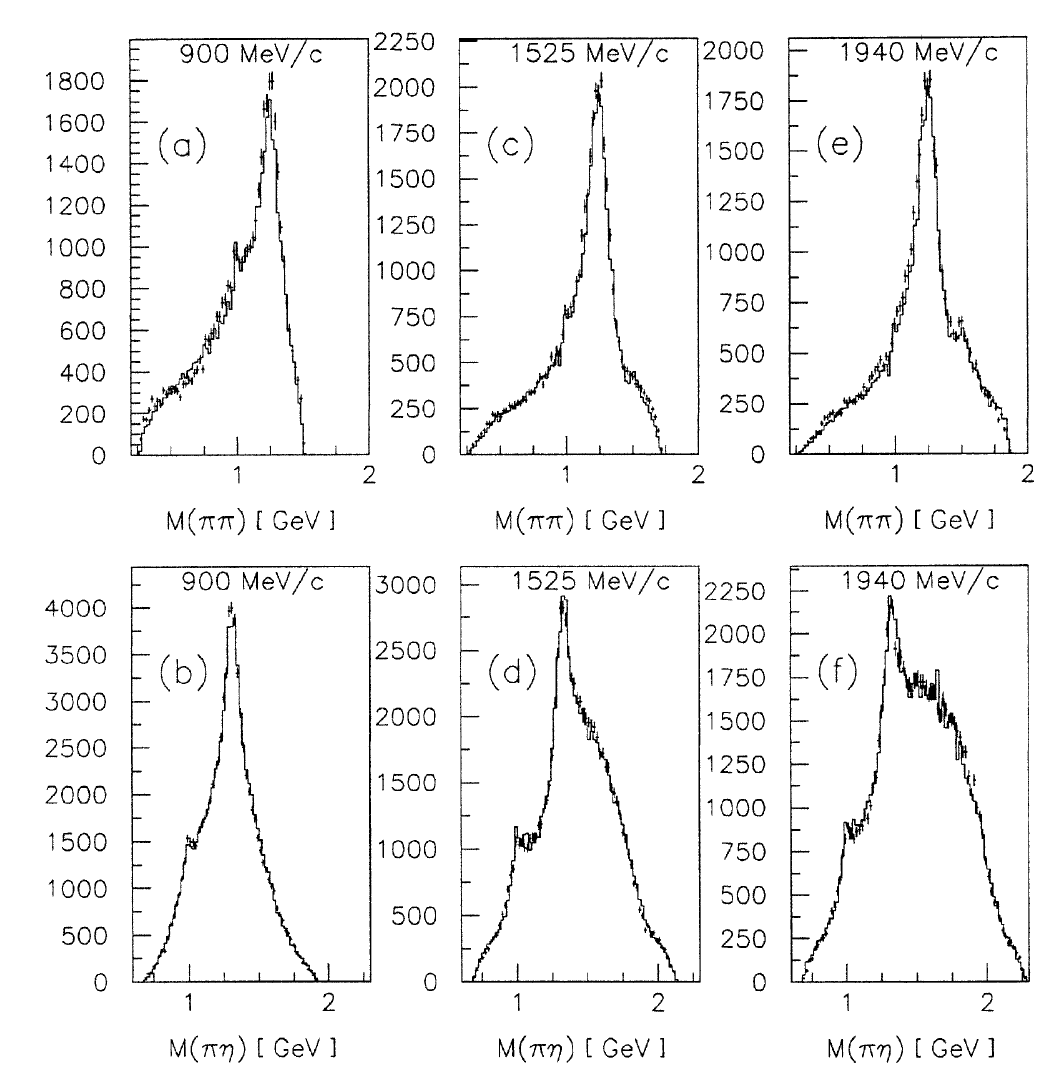}}
\caption{Mass projections on to $M(\pi \pi )$ and $M(\pi \eta )$ for
data at (a) and (b) 900 MeV/c, (c) and (d) 1525 MeV/c, (e) and (f) 1940 MeV/c;
histograms show the result of the fit.
Data are uncorrected for (small) variations of acceptance; these are included
in the maximum likelihood fit. }
\end{figure}

\begin{figure}
\centerline{\includegraphics[width=0.8\textwidth]{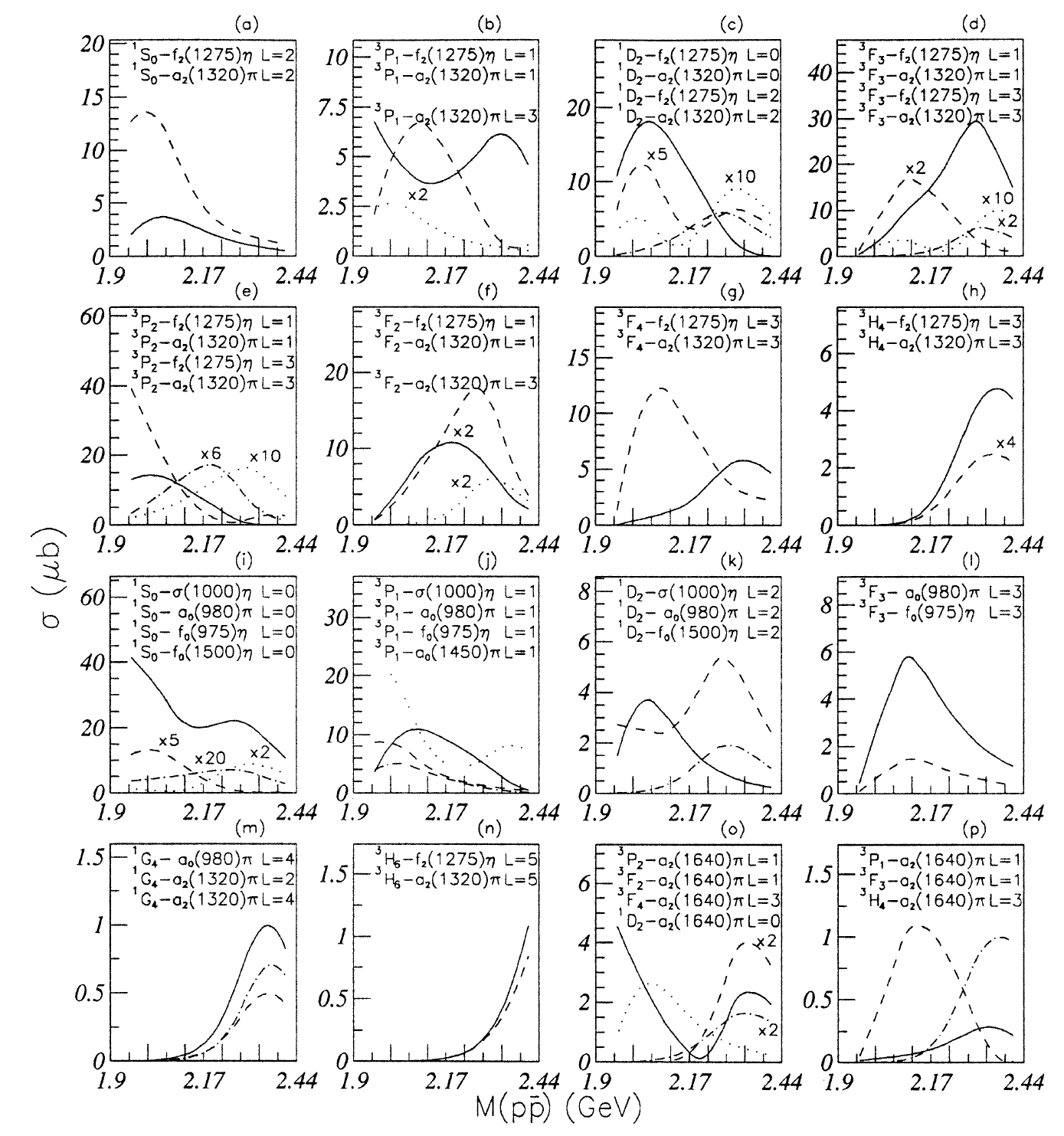}}
\caption{Intensities fitted to individual partial waves, and
corrected for all $\eta$ and $\pi ^0 $ decays.
The titles in
each panel indicate in descending order full, dashed, chain and
dotted curves; $\sigma (1000)$ denotes the broad $\sigma$ ($f_0(400-1200)$)
in the $\pi \pi$ amplitude.
Numerical values indicate factors by which small intensities
have been scaled so as to make them visible.}
\end{figure}

\begin{figure}
\centerline{\includegraphics[width=0.8\textwidth]{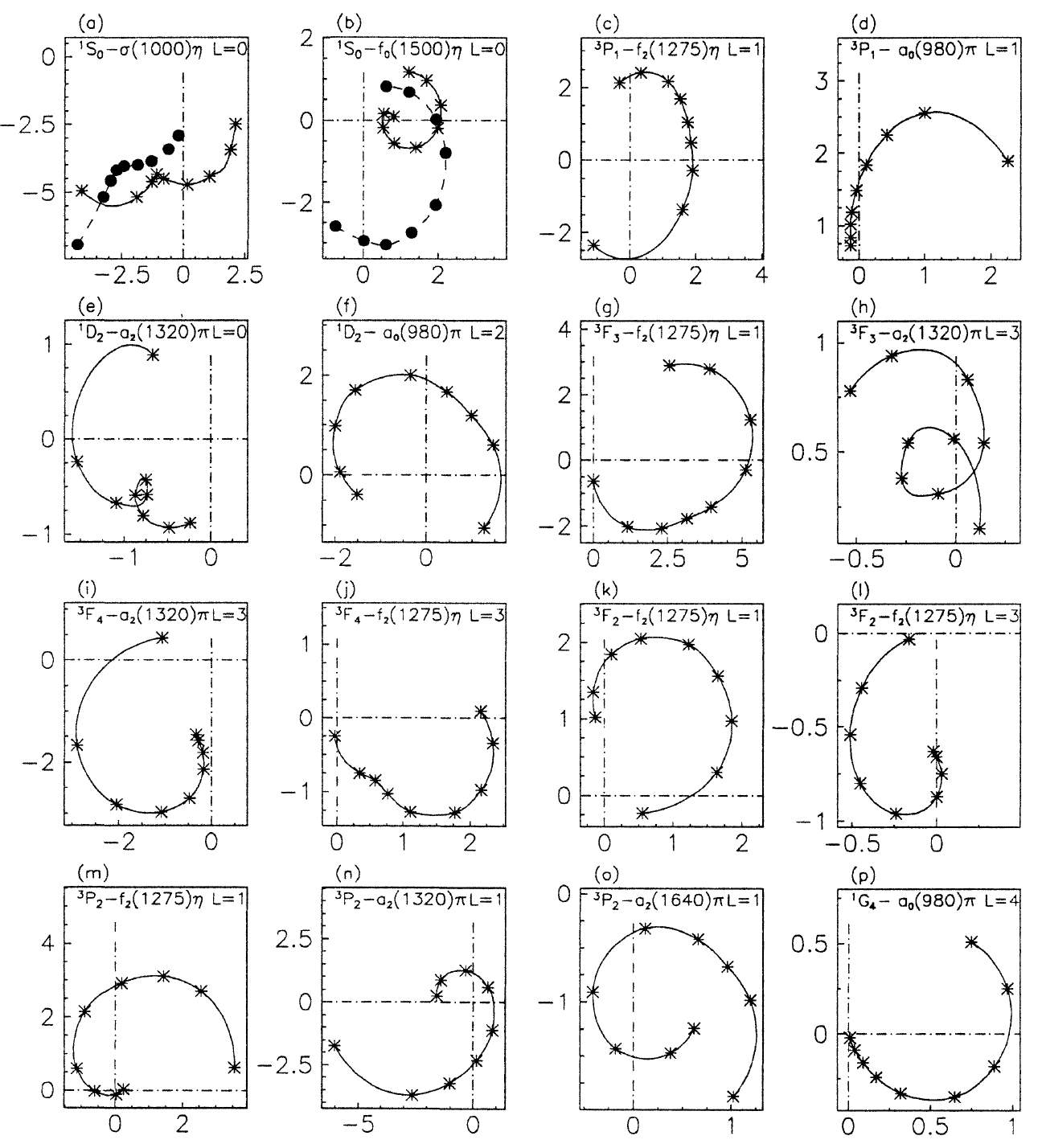}}
\caption{Argand diagrams for prominent partial waves; for (a) and (b), dashed
curves show the effect of removing the lower $0^-$ resonance.
Crosses mark beam momenta of Table 1; in all cases, the
amplitude moves anti-clockwise with increasing beam momentum.}
\end{figure}

\begin{figure}
\centerline{\includegraphics[width=0.8\textwidth]{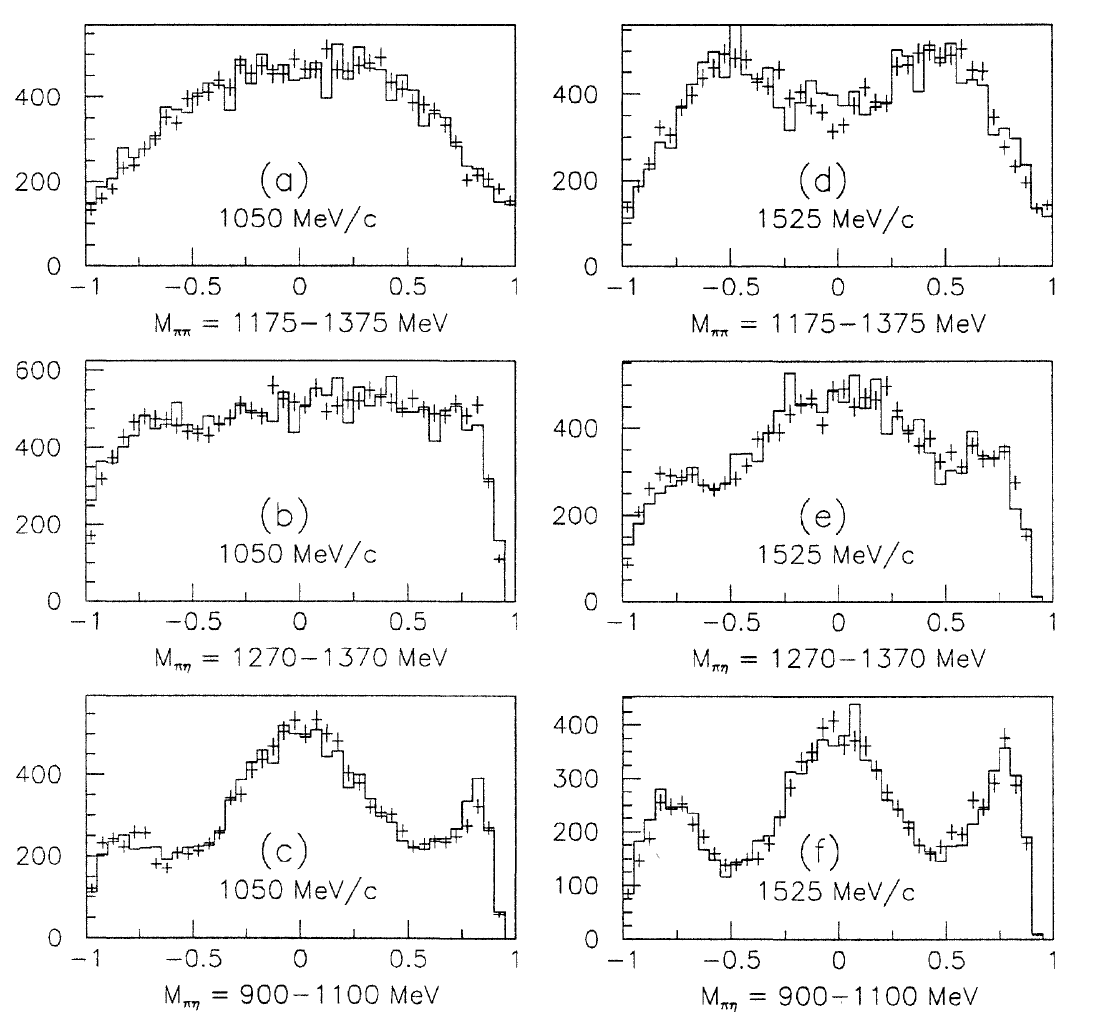}}
\caption{Centre of mass angular distributions for production of
(a) $f_2(1270)\eta$, (b) $a_2(1320)\pi$, (c) $a_0(980)\pi$ at 1050 MeV/c;
(d)-(f) show corresponding angular distributions at 1525 MeV/c.
Histograms show the fit. Data and fit are uncorrected for angular acceptance
of the detector; this acceptance is included in the maximum likelihood fit.}
\end{figure}

\begin{figure}
\centerline{\includegraphics[width=0.8\textwidth]{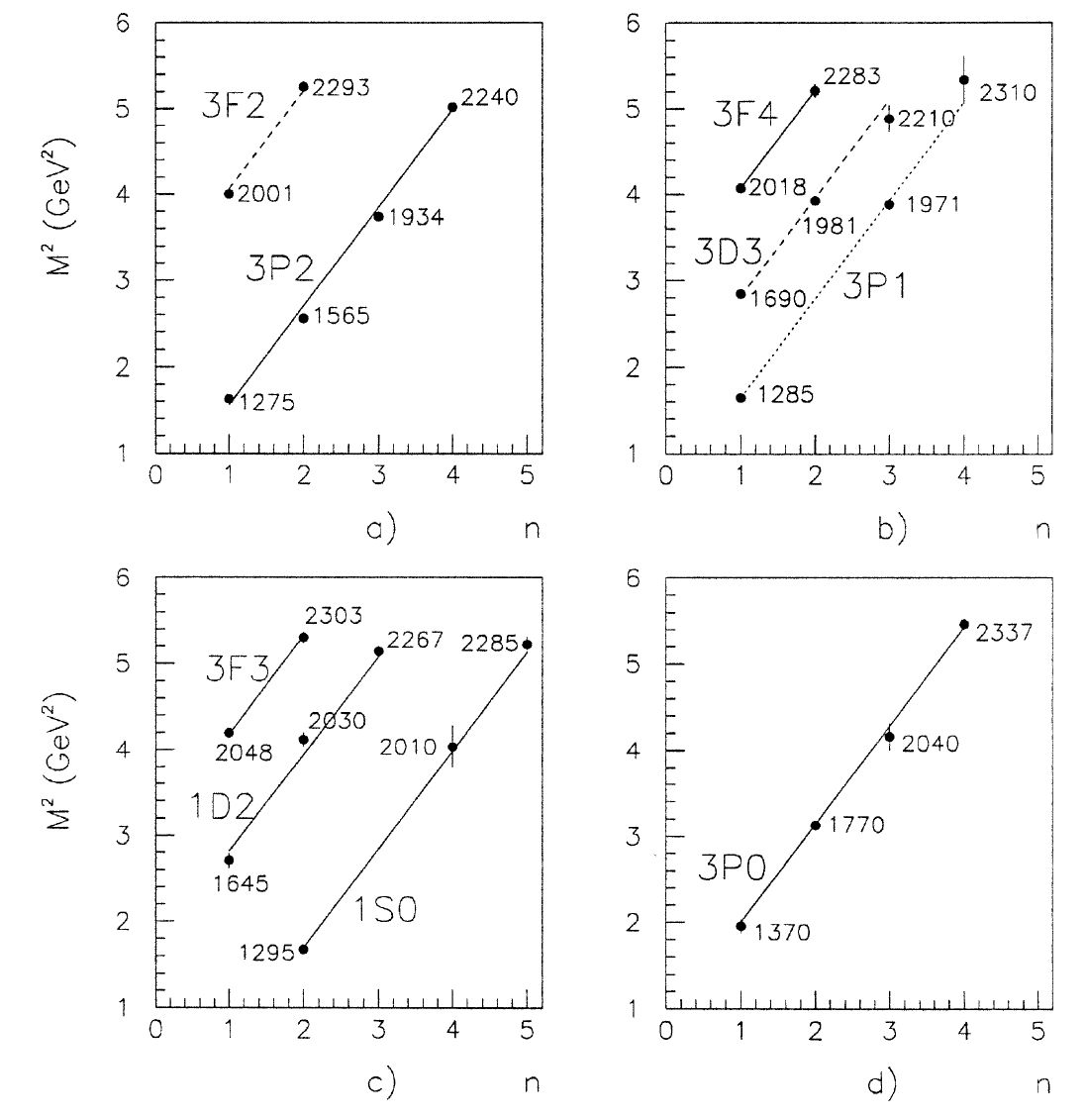}}
\caption{Plots of $M^2$ v. radial excitation number, $n$. In (b),
$I = 1$ $^3D_3$ states are shown. Straight line
trajectories are drawn in all cases with a slope of 1.143 GeV$^{-2}$. }
\end{figure}

\end {document}